\documentclass[aps,prl,preprint,groupedaddress]{revtex4-1}

\usepackage{graphicx}
\usepackage{dcolumn}
\usepackage{bm}
\usepackage{color}

\usepackage{comment}
\usepackage[geometry]{ifsym}
\usepackage{booktabs}
\begin{document}


\title{Deformation of  semicircle law  \\ for  correlated time series and Phase transition
}



\author{Masato Hisakado}
\email{hisakadom@yahoo.co.jp} 
\affiliation{
* Nomura Holdings, Inc., Otemachi 2-2-2, Chiyoda-ku, Tokyo 100-8130, Japan} 
\author{Takuya Kaneko}
\email{tkaneko@icu.ac.jp}
\affiliation{
\dag 
International Christian  University \\
Osawa 3-10-2, Mitaka, Tokyo 181-8585, Japan}


\date{\today}

\begin{abstract}

We study Wigner-type random matrices constructed from financial time series with temporal correlations. 
We characterize the deformed spectral law through its moments and observe behavior consistent with a modified semicircle law.
We  apply our framework to  several financial time series and observe  deviations from the semicircle law in specific foreign exchange markets. 
The difference from the semicircle law  for the financial  time series  depends on the temporal correlation of financial time series. 
We provide a moment analysis for both exponential and power law correlation structures and show that the fourth moment increases with the strength of correlations.
In the case of power law decay, a transition emerges between regimes with finite and divergent higher-order moments.  
Finally, numerical simulations support the analytical predictions and reveal finite-size scaling behavior near the  transition.

\hspace{0cm}
\vspace{1cm}
\end{abstract}


\maketitle

\section{I. Introduction}

Random matrix theory (RMT) provides a universal framework for describing the spectral behavior of large complex systems \cite{Meh}. Depending on the structure of the underlying random matrix, distinct limiting eigenvalue distributions arise. For instance, the Marchenko–Pastur distribution (MPD) governs the spectrum of Wishart matrices, while the Wigner semicircle law characterizes the spectrum of symmetric random matrices with independent entries. Deviations from these classical laws often signal additional structures such as correlations or heavy-tailed statistics, and have been widely studied in physics, mathematics, and applications ranging from nuclear spectra to network theory \cite{Pot,Pot3,Pot2,Fra,RM,AI}.

In this article,  
we construct   a Wigner  matrix  using the financial  time series.
 Several  financial time series exhibit temporal correlations.
When  the time series   are uncorrelated, the distribution of the eigenvalues of the   matrix  converges to  the semicircle law.
The exponential decay corresponds to a short memory,  and the power decay corresponds to
intermediate or long memories  \cite{Long}.
In  financial time series, power decays  are sometimes   observed 
as   fractional Brownian motion (fBm),  which includes 
both long and short memories \cite{fbm1, fbm2, fbm3, fbm4}.
The deformation of the semicircle law  is also  observed  in  the Wigner-L\`{e}vy matrix \cite{B1,B2,LD,tao,tao2} and has  a longer  tail and a higher peak.
When the matrix entries are Lévy-distributed and $N\rightarrow \infty$, the scaled eigenvalue density converges to a limiting density that differs from the semicircle law.
The  limiting density   depends on the degree of  freedom of the t distribution.
Financial returns are known to exhibit fat-tailed distributions \cite{man}. 
The financial time series have both properties,  temporal correlation and fat tail.
Regarding  the financial time series, we observed  a difference from the   semicircle law.
We test the null hypothesis that the time series is normal i.i.d. using the Wigner matrix and  observe the difference from the semicircle law for  the FX time series.
We show the origin of the difference is temporal correlation.

Moreover, we study the effects of the  temporal correlations of  random variables.
The temporal correlation is the exponential decay and power decay.
When there are temporal correlations,
the eigenvalues   of the Wigner matrix  are consistent with  the deformed  semicircle law.
We  show  that    fourth  moment increases as the temporal correlation increases.
Hence, as the correlation increases, the distribution has a fatter tail and a higher peak.
In  the power decay case,
we    observe   a  phase transition.
There are  finite fourth moment  and  infinite fourth moment phases.
When  $\gamma>1/2$, which is the power index of the temporal correlation, the fourth moment of the distribution  and the largest  eigenvalue are  finite.
On the other hand,   when $\gamma\leq 1/2$, the  fourth moment and the largest  eigenvalue are  infinite.
In fact,     phenomena  such as   phase transition   depend on the  temporal correlation \cite{Hisakado6, Hisakado7}. 
 
In previous studies, the effect of temporal correlations on Wishart matrices has been shown to deform the Marchenko–Pastur distribution, leading to heavier tails and the emergence of phase transition \cite{MP, Hisakado5}. 

In these systems, the transition is associated with the divergence of low-order moments and can be characterized by a critical exponent governing finite-size scaling behavior.

Despite these advances, an important question remains open: are correlation-induced spectral transitions universal across different random matrix ensembles, or do they depend on the structural properties of the matrices? In particular, it is not clear whether the mechanism identified in the Wishart ensemble extends to Wigner-type matrices, which differ fundamentally in symmetry and construction.

In this work, we address this question by studying Wigner-type random matrices constructed from time series with temporal correlations. We show that correlations deform the semicircle law and give rise to a phase transition in the case of power-law decay. Unlike the Wishart case, where the transition is governed by the second moment, the present system is controlled by the fourth moment, which increases monotonically with the strength of correlations. Furthermore, we find that the critical exponent associated with finite-size scaling differs from that of the Wishart ensemble.

These results indicate that correlation-induced spectral transitions are not universal, but instead depend sensitively on the underlying matrix structure. In other words, Wigner and Wishart ensembles belong to distinct universality classes with respect to correlation effects.

This distinction can be understood qualitatively from the difference in their construction: Wishart matrices are formed as products of data matrices, while Wigner matrices consist of symmetric entries, leading to different combinatorial structures in moment expansions.

In the Wishart ensemble, the leading correlation contribution appears at the level of the second moment because the matrix is formed as a product, and the moment expansion involves pair contractions of off-diagonal entries at order $N^2$. In the Wigner case, the symmetric construction and the folding procedure shift the leading sensitive moment to the fourth order, where non-trivial combinatorial factors — specifically the fraction of correlated pairs crossing the diagonal — first introduce correlation-dependent terms.

To validate the theoretical predictions, we perform numerical simulations using both synthetic data, including fractional Brownian motion, and empirical financial time series. The results support the moment-based analysis and reveal finite-size scaling behavior consistent with the proposed framework.

The remainder of this paper is organized as follows.
In Section II, we introduce  the time series  and the creation of the Wigner matrix.
In Section III, we 
discuss the distribution of the  deformed 
semicircle law.
In Section IV, we apply our framework to the financial time series and compare to the semicircle law.
In Section V, numerical simulations are performed to confirm the deformed semicircle law by temporal correlation and fat tail of time series.
In Section VI, we study the phase transition of  the deformed semicircle law.
Finally, the conclusions are presented in Section VII.

\section{II. Time
series and the creation of the Wigner matrix}

In this section,  we  introduce  the Wigner     matrix with  correlation. 
We consider  the time series of a stochastic process as follows,  
$S_t$  are  the   variables at  time $t$.
In the case of  financial data,
we use the historical data of the return $r_t$ as $S_t$.
The return is defined using the market price, $p_t$ as
\[
r_t=\ln p_{t}-\ln p_{t-1}.
\]
Here, we  set the normalization,
\[\mbox{E}(A_t)=0,
\]
and 
\[\mbox{V}(A_t)=1.
\]
To introduce the temporal  correlation, let $\{A_t,1\le t \le T\}$ be the time series of  stochastic
  variables of a correlated normal distribution with the following
 $T\times T$ correlation matrix,
\begin{equation}
D_{T-1}=\left(
    \begin{array}{cccc}
     1 &  d_1 &\cdots& d_{T-1} \\
   d_1      &  1 & \ddots&\vdots \\
 \vdots & \ddots   &1&d_1 \\
    d_{T-1}&  \cdots &  d_1  & 1  \\
    \end{array}
  \right),
  \label{matrix}
\end{equation}
which parameterizes   the time series.
Here, the temporal correlation function, $d_t$, is defined as the correlation between $A_i$ and $A_{i+t}$ such that
\begin{equation}
d_t= 
\mbox{Cov}(A_i,A_{t+i}),
\label{Cor}
\end{equation} 
for any $i$.
In this study, we  consider the exponential decay and power decay cases.

Note that $T=N^2$.
The time series data $A_t$, $t=1,\cdots, T$   are folded $N$ times, and unnecessary parts are removed.
In the matrix form, we can write  the $N\times N$ matrix
form
\begin{eqnarray}
S&=&S_{ij}=
\left(
    \begin{array}{ccccc}
     A_1 &  A_{2}&A_{3}  &\cdots& A_{N} \\
   A_2    &  A_{N+2} & A_{N+3}&\cdots& A_{2N} \\
   A_{3} &A_{N+3}& A_{2N+2}& \cdots           & A_{3N} \\
 \vdots & \vdots & \ddots &\cdots &\vdots\\
    A_N& A_{2N}& A_{3N} &  \cdots &A_{N^2} \\
    \end{array}
  \right).
  \label{EX}
\end{eqnarray}
This construction ensures that the matrix elements inherit the temporal correlation structure of the time series.
Note that   we consider the case $N>>1$.
Hence,
\[
d_i= 
\mbox{Cov}(A_i,A_{i+j})=0,
\]
$j\geq N$.

In another  definition,
matrix $S$ has the following  correlation,
\begin{equation}
{\rm Cov}(S_{ij}, S_{ij+t})=d_t,
\end{equation}
where $i\geq j $.

\section{III. Deformed  semicircle law distribution and effects of temporal correlation}

In this section,  we calculate the moments of the deformed semicircle law for the effects of the temporal correlation.
Therefore, we assume the normal distribution with temporal correlation.
Here, we consider the case $d_{i}\rightarrow 0$, when $i>>1$.
This  means  that the temporal correlation decays as time goes by.
We  calculate the $k$-th moment of  the eigenvalue distribution of the semicircle law, $S$,  
\begin{equation}
\mu_k=\frac{1}{N^{(k+2)/2}}<\sum_{j=1}^N (x_j)^k>=
\frac{1}{N}<{\rm Tr}(S^k)>,
\end{equation}
where $x_j$ is the eigenvalue of $S$ and
$<>$ means the ensemble average.
$S$ is the symmetric, and the odd  moments  vanish.

\subsubsection{i. Second moment}
\begin{eqnarray}
    \mu_2&=&\frac{1}{N^{2}}\sum_{j=1}^N \sum_{l=1}^N
    <(S_{j l}S_{l j})>
    \nonumber \\
   &=&\frac{1}{N^2} (\sum_{j=1}^N <(S_{jj})^2> +2\sum_{j<l}^N <(S_{jl})^2>)=1, 
\label{av}
\end{eqnarray}
in the limit of $N \rightarrow \infty$.
It should be noted  that it does not depend on the type of  correlation decay.
The mean of the distribution that  does not depend on the  correlation is 1.
Therefore, we obtain 
\begin{equation}
 \sum_{i=1}^N x_i^2=N,
 \label{sq}
\end{equation}
where $x_i$ is the  eigenvalue,
$x_1\geq x_2 \geq \cdots \geq x_N$.
Therefore, $x_1$ is the largest eigenvalue. 

\subsubsection{ii. Fourth  moment}
\begin{eqnarray}
    \mu_4&=&\frac{1}{N^3}\sum_{j=1}^N\sum_{l 
 =1}^N\sum_{m =1}^N
    \sum_{n =1}^N< S_{j l} S_{l m}S_{m n} S_{n j}> \nonumber \\
    &=&
    2+\frac{4}{3} \sum_{i=1} d_i^2,
\label{mu}
\end{eqnarray}
in the limit of $N \rightarrow\infty$.
We assume $d_N\sim0$ in this limit, because 
\[
<S_{ij},S_{kl}>=0,
\]
when both $i\neq k$ and $j\neq l$ in the limit $N\rightarrow \infty$.
The calculation of the moment of the  semicircle law  is presented in Appendix A.

The first term  in  Eq.(\ref{mu}) corresponds to    the semicircle law.
The second  term refers to  the   correlation.
The second term of Eq.(\ref{mu}) increases as the  correlation increases.
Note that it does not depend on the sign of the  correlations.
The deformed semicircle law has a longer tail  and  a higher central peak than  the semicircle law.
If the  fourth  moment is finite, then  the    largest eigenvalue  does not diverge.
However, the  fourth moment is infinite, and the  largest eigenvalue is infinite.
We discuss  phenomena  such as  phase transition in  the subsection for the power decay case.

\subsubsection{iii. Sixth  moment}
\begin{eqnarray}
    \mu_6&=&\frac{1}{N^4}\sum_{j=1}^N\sum_{l 
 =1}^N\sum_{m =1}^N
 \sum_{n 
 =1}^N\sum_{o =1}^N
    \sum_{p =1}^N< S_{j l} S_{l m}S_{m n} S_{n o}S_{o p} S_{p j}> \nonumber \\
    &=&
    5+8 \sum_{i=1} d_i^2+\frac{8}{3} \sum_{i=1} d_i^4,
\label{mu6}
\end{eqnarray}
in the limit of $N \rightarrow\infty$.

\subsection{A. Exponential decay case}
We consider the exponential decay case, 
$d_{i}=\mbox{Cov}(A_s,A_{i+s})=r^i,0\le
r\le 1$ and calculate the moments.
$r$ is the effect at $t$ from the past at  $t-1$ and 
a commonly used  temporal correlation for financial time series. 
Here, we set 
\begin{equation}
A_{t+1}=rA_t+\sqrt{1-r^2}\xi_t,
\label{A}
\end{equation}
where $\xi_t$ is i.i.d. and we  obtain  the exponential decay  to create a time series with  exponential decay, $<A_{t+1},A_t>=r$.

The fourth  moment is 
\begin{equation}
    \mu_4
    =
    2+\frac{4}{3}\frac{r^2}{1-r^2},
    \label{expmu}
\end{equation}
in the limit of $N\rightarrow \infty$.
The first term is for  the  semicircle law.
The second  term is for the deformation for   the correlation.
Note that  the fourth moment  is finite, because $r<1$.

The sixth   moment is 
\begin{equation}
    \mu_6
    =
    5+8\frac{r^2}{1-r^2} + \frac{8}{3}\frac{r^4}{(1-r^2)^2},
    \label{expmu2}
\end{equation}
in the limit of $N\rightarrow \infty$.
Note that  the  sixth  moment  is finite value, because $r<1$.
Higher moments   are also finite, because the higher moments  are polynomial of $r^2/(1-r^2)$ which is finite for the exponential decay case.

\subsection{B. Power decay case}

\subsubsection{i.Fourth moment}
In this section, we consider the case
 of power decay, $d_i=\mbox{Cov}(A_s,A_{i+s})=1/(i+1)^{\gamma}$ where
$\gamma$ is the power index.
The fourth moment converges to the finite value, when $\gamma>1/2$, 
\begin{eqnarray}
    \mu_4&=&
    2+\frac{4}{3}\sum_{i=1}^{\infty}\frac{1}{(i+1)^{2\gamma}}
    \nonumber \\
    &<&
  2+\frac{4}{3}\int_{1}^{\infty}\frac{1}{(x+1)^{2\gamma}}dx
    \nonumber \\   
    &=&
    2+\frac{2^{3-2\gamma}}{3(2\gamma-1)}, 
    \label{powmu}
\end{eqnarray}
in the limit of $N\rightarrow \infty$.
Conversely, $\gamma\leq 1/2$, 
\begin{eqnarray}
    \mu_4&=&
    2+\frac{4}{3}\sum_{i=1}^{\infty}\frac{1}{(i+1)^{2\gamma}}
    \nonumber \\
    &>&
 2+\frac{4}{3}\int_{2}^{\infty}\frac{1}{(x+1)^{2\gamma}}dx
 \nonumber \\   
    &\sim&
    \lim_{x\rightarrow \infty}
   \frac{4}{3(1-2\gamma)}x^{-2\gamma+1}. 
\end{eqnarray}
The divergence of the fourth moment suggests a transition at
$\gamma_c=1/2$.

The sixth   moment is 
\begin{equation}
    \mu_6
    =
    5+8 \sum_{i=1}^{\infty}\frac{1}{(i+1)^{2\gamma}}  + 
    \frac{8}{3}\sum_{i=1}^{\infty}\frac{1}{(i+1)^{4\gamma}} ,
    \label{powmu2}
\end{equation}
in the limit of $N\rightarrow \infty$.
Note that  the   sixth  moment becomes infinite 
and 
the transition point is $\gamma_c=1/2$ because of the second term.
Higher moments over  moments  are also finite, because the higher moments  includes the  polynomial of 
$\sum_{i=1}^{\infty}1/(i+1)^{2\gamma}$.

Higher moments  are  also polynomial of $\sum 1/(i+1)^{2\gamma}$.
Therefore, when $\gamma>1/2$, all moments are finite.
On the other hand, $\gamma<1/2$, all moments are infinite.

In this section we assume the case $d_{i}\rightarrow 0$, when $i>>N$.
We calculate the  finite  size correction.
we define the effect $\Delta \mu_4$ which is the finite size tail.
In the case of power decay case,
\begin{eqnarray}
\Delta \mu_4 &\sim& \sum_{i}d_{iN}^2\sim N^{1-4 \gamma}.
\end{eqnarray}
Due to the folding construction, for the upper(lower) triangle matrix, the correlations between entries in different columns(rows) involve index separations of order $N$, effectively generating long-range contributions even for finite index differences.
When $\gamma>1/4$, $\Delta \mu_4$ converges to 0.
However, when $\gamma<1/4$,  $\Delta \mu_4$   diverges.
There are two distinct scaling regimes. The bulk contribution diverges at $\gamma= 1/2$, while the finite-size correction induced by the folding structure exhibits a separate scaling transition at $\gamma=1/4$.
Although the finite-size correction diverges for $\gamma<1/4$, it remains subleading compared to the bulk contribution, and therefore does not affect the leading-order scaling of the fourth moment.
For $N=1024$ and $\gamma=0.6$ (near the critical point), the finite-size correction $\Delta \mu_4 \sim  0.0006$, which is much less than  the leading-order contribution $\mu_4=2.05$ (Table III).

In this section we discussed using the moment calculations.
While the moment calculations provide a useful characterization of the spectral distribution, they do not by themselves guarantee weak convergence. A rigorous proof would require tools such as the Stieltjes transform, which is beyond the scope of the present work.

\section{IV. Application to financial data}


Here, we calculate  the eigenvalue distribution  of the Wigner matrix
and compare the moments  with  the  semicircle law  
 for 25 financial time series,  which includes
 Crypt assets, foreign exchange, commodities, government bonds, and stock indices \cite{B}. 
We use    time series of one product and confirm the difference from the semicircle law.
We report   the $5\%$  and  $95\%$  quantile of normal  i.i.d.  in Table \ref{5p} as  a  test. 
We repeat the numerical simulation of the moments for normal i.i.d. 10000 times for $N=64$, 5000 times for $N=128$, 5000 times for $N=256$, and $1000$ times for $N=512$.
The larger number of pairs indicates 5 percentile largest moments, whereas the smaller number indicates 95 percentile largest number of  repeats.
For   the null hypothesis, the time series is normal i.i.d.
Outside of this range  the null hypothesis is rejected. 
\begin{table}[ht]
\centering
\begin{tabular}{c|cc|cc|cc|cc|cc|cc}
\hline
N $/$ $\mu_i$ & \multicolumn{2}{c|}{$\mu_1$} & \multicolumn{2}{c|}{$\mu_2$} & \multicolumn{2}{c|}{$\mu_3$} & \multicolumn{2}{c|}{$\mu_4$} & \multicolumn{2}{c|}{$\mu_5$} & \multicolumn{2}{c}{$\mu_6$} \\
\hline
512 & -0.0032 & 0.0031 & 0.9934 & 1.0062 & -0.0125 & 0.0116 & 1.974 & 2.0286 & -0.0531 & 0.0483 & 4.8952 & 5.1255 \\
256 & -0.0066 & 0.0064 & 0.987 & 1.013 & -0.0246 & 0.0244 & 1.9503 & 2.0589 & -0.1016 & 0.1019 & 4.8053 & 5.2578 \\
128 & -0.0128 & 0.0128 & 0.9743 & 1.0251 & -0.0502 & 0.0498 & 1.8992 & 2.1157 & -0.2132 & 0.2091 & 4.6202 & 5.5042 \\
64 & -0.0255 & 0.0255 & 0.9499 & 1.0508 & -0.1002 & 0.0993 & 1.8033 & 2.2381 & -0.4196 & 0.4192 & 4.2665 & 6.0532 \\
\hline
\end{tabular}
\caption{The table is  the matrix size $N$ and   the $5\%$  and  $95\%$  quantile   of the numerical simulation of  the moments for normal i.i.d.  Outside of this range  is the  rejection of the normal i.i.d. 
 The quantile bands  are constructed under the i.i.d. Gaussian assumption and serve as a reference benchmark. Under correlated or heavy-tailed data, the sampling distribution of moments may differ, and therefore the results should be interpreted as indicative rather than as formal hypothesis tests.}
\label{5p}
\end{table}

\begin{figure}[htbp]
\begin{tabular}{ccc}
\begin{minipage}{0.33\hsize}
\begin{center}
 \includegraphics[width=5.cm]{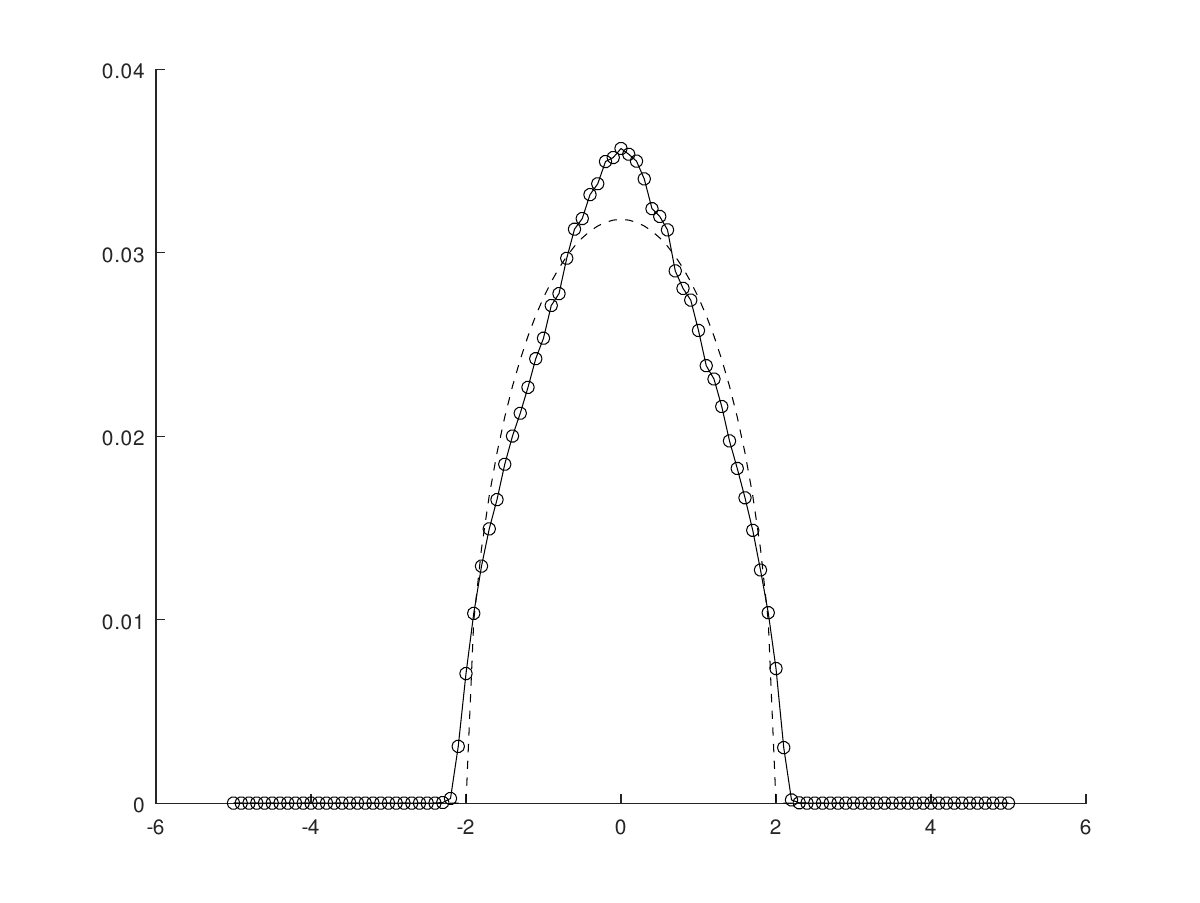} 
\hspace{1.6cm} (a) No.7
\end{center}
\end{minipage}
 &
\begin{minipage}{0.33\hsize}
\begin{center}
 \includegraphics[width=5.cm]{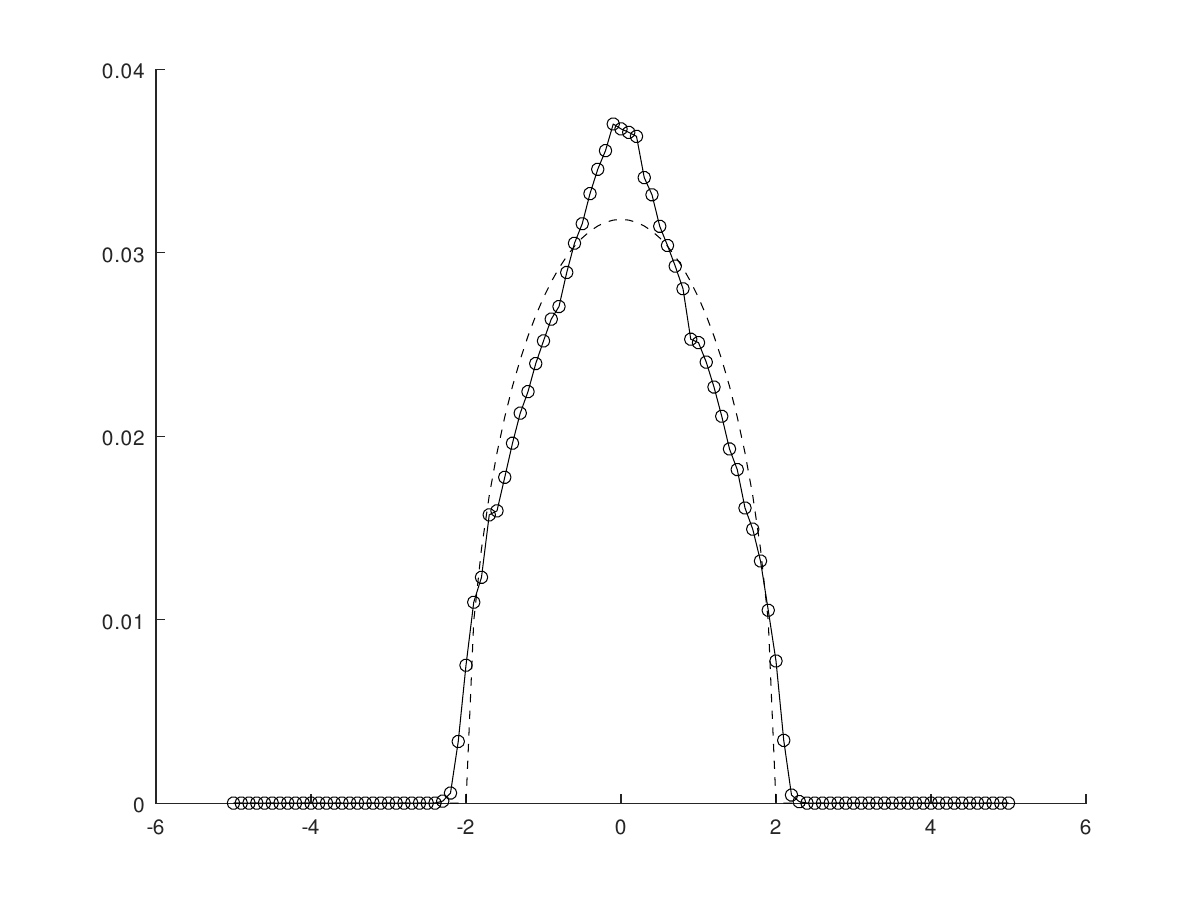} 
\hspace{1.6cm} (b) No.9
\end{center}
\end{minipage}
 &
 \begin{minipage}{0.33\hsize}
\begin{center}
 \includegraphics[width=5.cm]{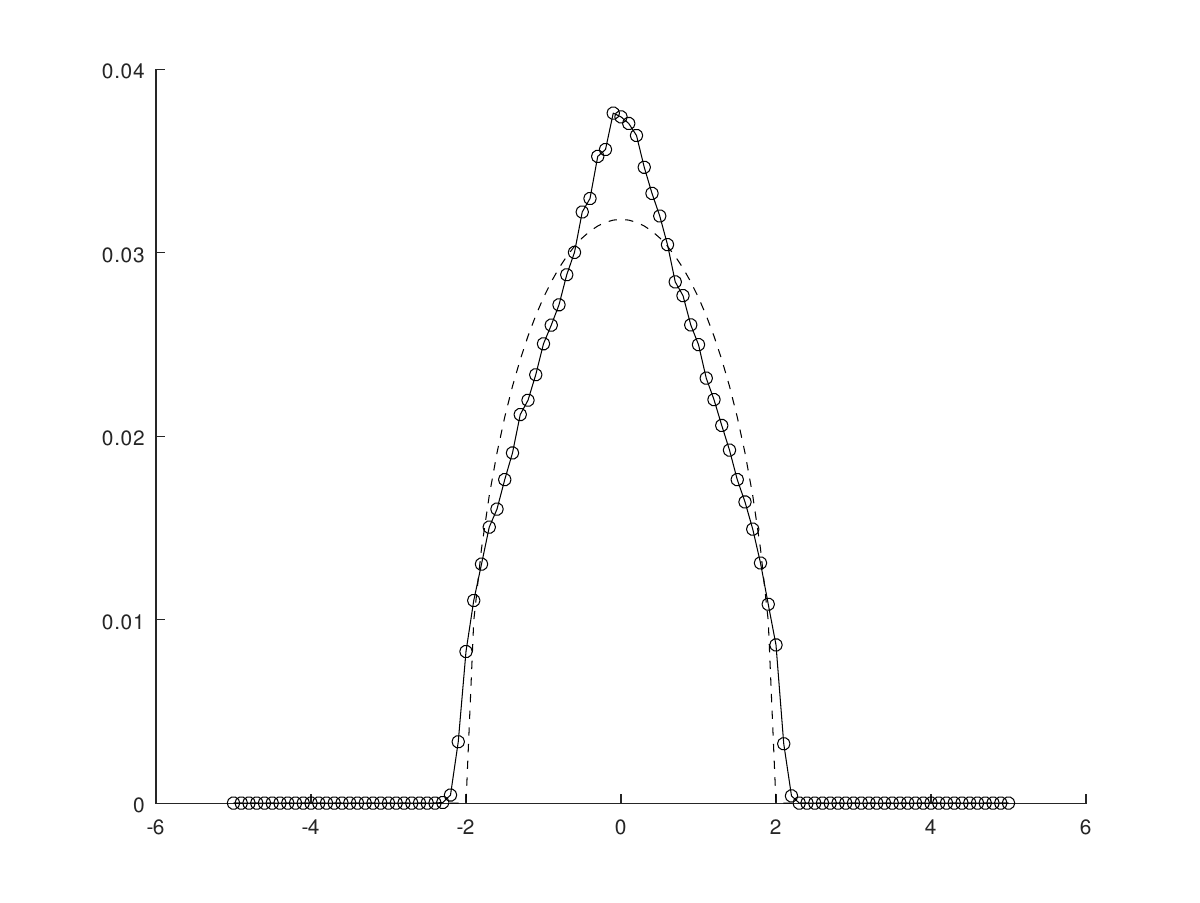} 
\hspace{1.6cm} (c) No.12
\end{center}
\end{minipage} 
\\
\begin{minipage}{0.33\hsize}
\begin{center}
 \includegraphics[width=5.cm]{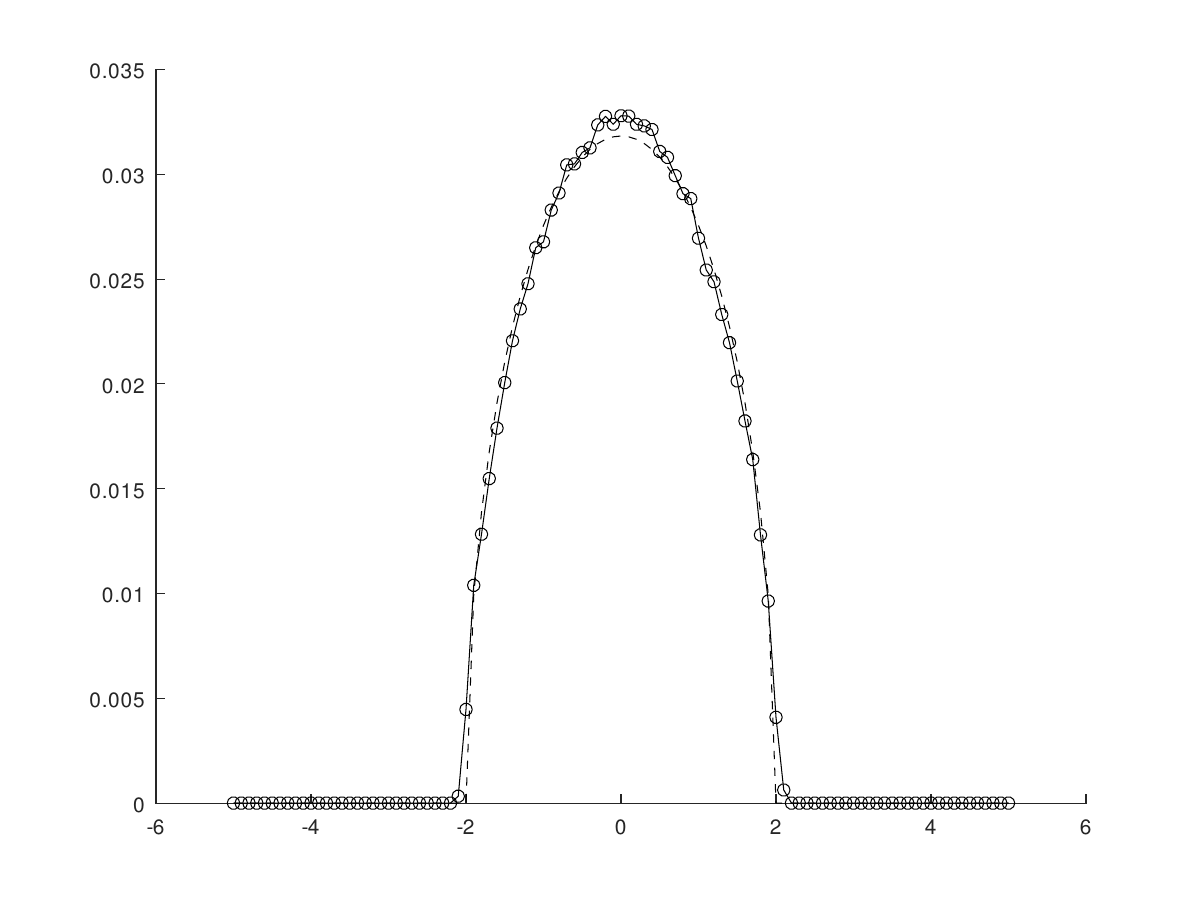} 
\hspace{1.6cm} (d) No.19
\end{center}
\end{minipage}
 &
\begin{minipage}{0.33\hsize}
\begin{center}
 \includegraphics[width=5.cm]{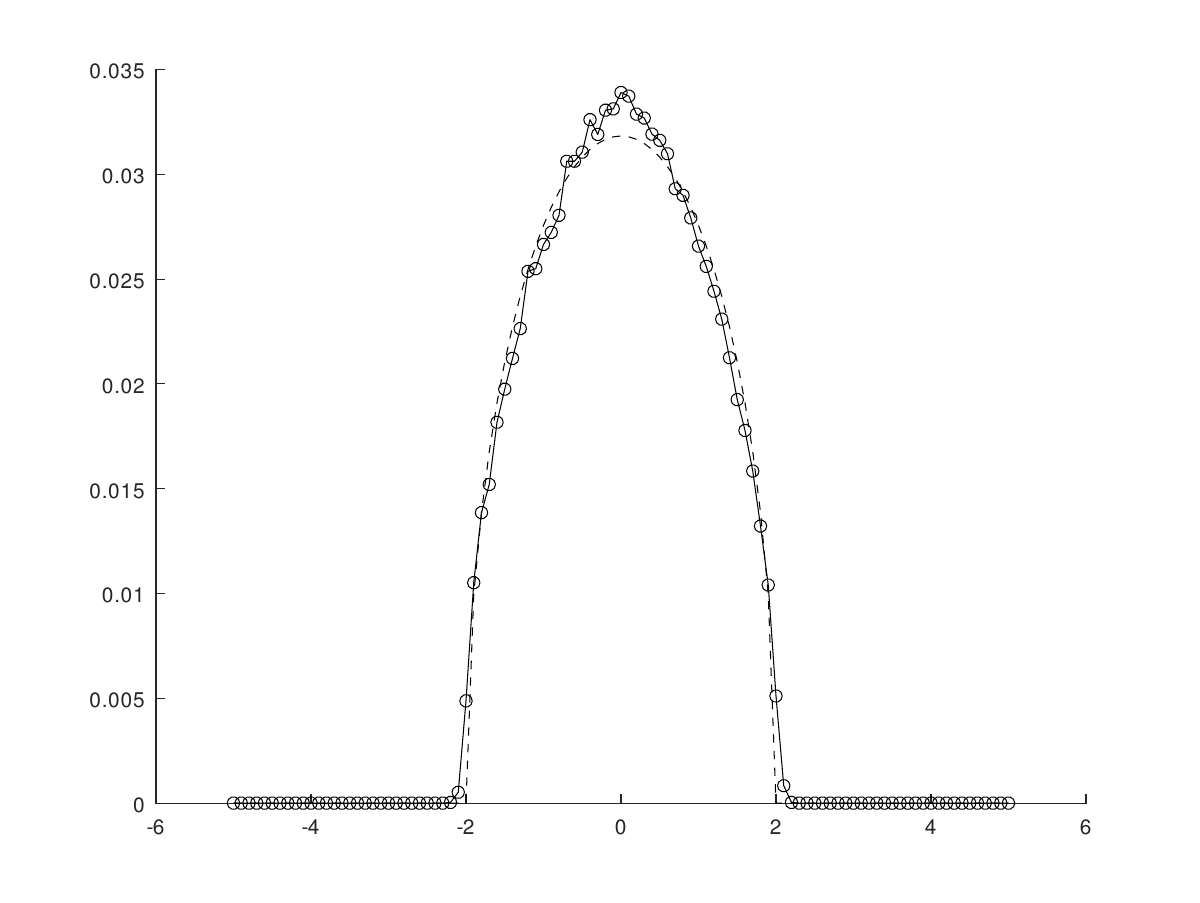} 
\hspace{1.6cm} (e) No.22
\end{center}
\end{minipage}
 &
 \begin{minipage}{0.33\hsize}
\begin{center}
 \includegraphics[width=5.cm]{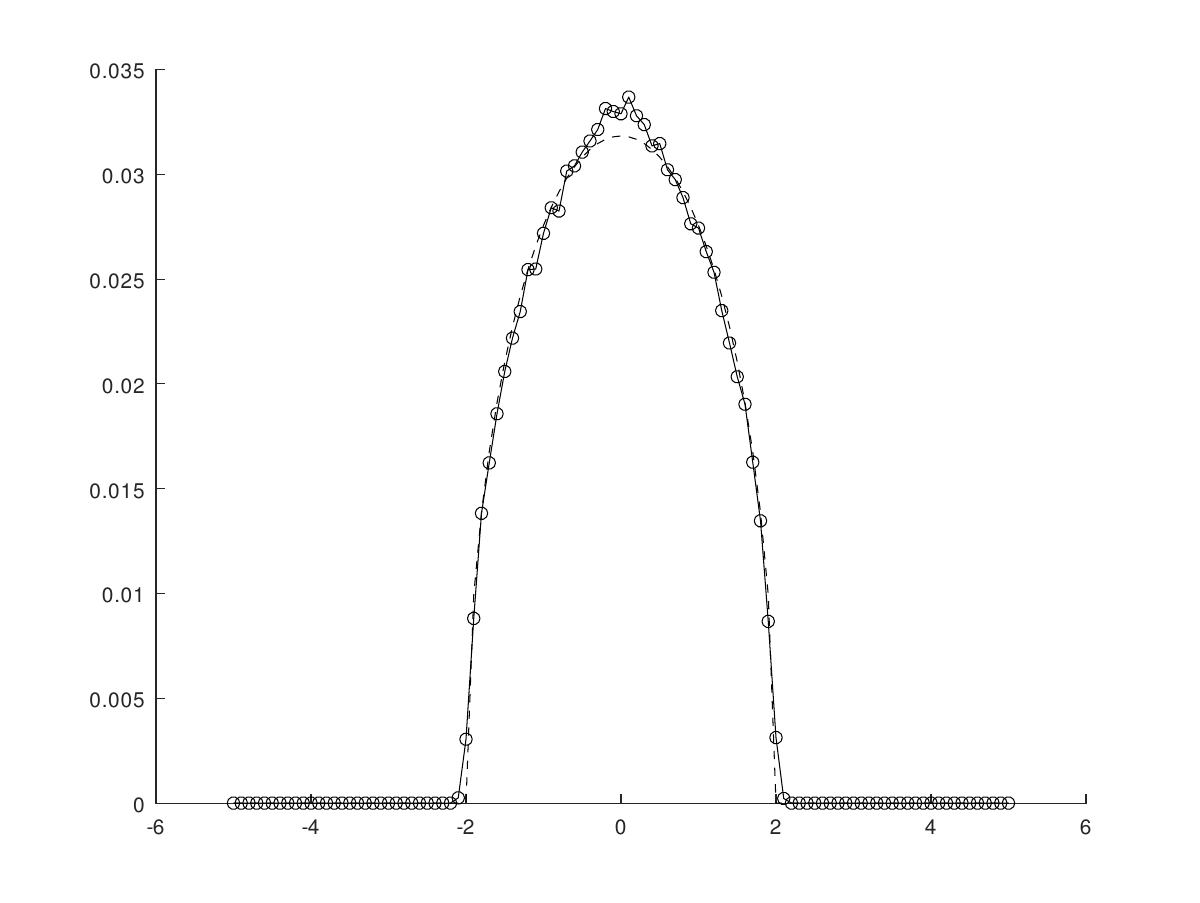} 
\hspace{1.6cm} (f) No.23
\end{center}
\end{minipage} 
\end{tabular}  
\caption{Plots of the histogram of  financial  time series:
(a) No.7 USD/CHF, (b) No.9 USD/CAD, (c) No.12 EUR/CHF,
(d) No.19 WTI, (e) No.22 XAU, (f) No.23 VIX.
The horizontal axis represents  the eigenvalues and the vertical axis 
represents the frequency.
The real line presents the distribution  of the time series  and the dotted line represents  the semicircle law. 
We can confirm that the FX distributions have a higher central peak than the semicircle law.
}
\label{MPD3}
\end{figure}

The moments of the time series are shown   in Table \ref{table3}.
The comparison to   the null hypothesis,  time series is the normal i.i.d. is in 
Table \ref{T}.
We can confirm that  the moments for the   time series fit well with the semicircle law except for  FX and some  crypt assets.
Particularly, No.7 USD/CHF, No.9 USD/CAD, No.12 EUR/CHF,  and  No.13 EUR/GBP do  not fit  the semicircle law well.
In Fig.\ref{MPD3} some  distributions are shown.
We can confirm  that the distribution of  (a)-(c) (No.7, 9,12) has  a higher peak the semicircle law  in the central and  extension of the support.
Excluding  FX, the distribution (d)-(f) (No.19,22,23) fit  the semicircle law well.
One of the reasons for   the fit to the semicircle law  is the temporal correlations \cite{Hisakado5}.

The  time series  of No.1 BTC/USD, No.7 USD/CHF, No.9 USD/CAD,  No.12 EUR/CHF,  and No.13  EUR/GBP have higher  temporal correlations for the shortest time lag  as Table \ref{5p}.
This result is consistent with  Table  \ref{T}.

For the five time series for which the null hypothesis was rejected, we calculate the contribution of the first term of temporal autocorrelation to the fourth moment in Table \ref{cont}.
The first term is important, because  of the decrease  of temporal correlation for these time series.
The contribution of the first term  for the fourth moment is 
\begin{equation}
\mbox{contribution to the fourth moment }=\frac{4/3d_1}{\mu_4-2},
\label{4}
\end{equation}
and
\begin{equation}
\mbox{contribution to the sixth  moment }=\frac{8 d_1^2+8/3d_1}{\mu_6-5},
\label{6}
\end{equation}
This suggests that the deviation in the fourth and sixth  moments is largely attributable to the effect of temporal correlation.
Financial time series exhibit both properties, temporal correlation and fat tail.
In the following sections we show the effects of the temporal correlation  and 
the effects of the fat tail.


\begin{table}[htbp]
\centering
\caption{Moments for the distribution of eigenvalues of  financial time series and 
the semicircle law. $\mu_i$ is the $i$-th moment of the semicircle law and Corr. is the temporal correlation of the shortest time lag in the financial time series. 
The average of moments  were obtained using the bootstrap method. 
The bootstrap samples were generated using overlapping windows (sampling with overlap) in order to preserve the temporal correlation structure of the original time series as much as possible.
The estimation of  statistical uncertainties of $\mu_4$ and $\mu_6$ by bootstrap method is in Appendix D.
}
\begin{tabular}{|c|l|c|r|r|r|r|r|}
\hline
No & Data & N & $\mu_2$ & $\mu_4$ & $\mu_6$ & Corr. \\
\hline
0 & Theory & - & 1 & 2 & 5 & - \\
1 & BTC/USD & 512 & 0.9945709163 & 2.0322554437 & 5.2461161423 & -0.39832 \\
2 & XRP/USD & 256 & 0.992749619 & 2.033477488 & 5.272570986 & -0.014449 \\
3 & ETH/USD & 128 & 0.986091805 & 1.992831612 & 5.088267897 & 0.0052927 \\
4 & EUR/USD & 128 & 0.9919604654 & 2.0754426072 & 5.5220097433 & -0.22908 \\
5 & USD/JPY & 128 & 0.9903583699 & 2.0592822343 & 5.4526831969 & -0.20065 \\
6 & GBP/USD & 128 & 0.9892206415 & 2.0447484197 & 5.3697075654 & -0.16762 \\
7 & USD/CHF & 128 & 0.99117701133 & 2.12874344704 & 5.86166248676 & -0.30042 \\
8 & AUD/USD & 128 & 0.990287312 & 2.064446283 & 5.4880243294 & -0.17972 \\
9 & USD/CAD & 128 & 0.9923546213 & 2.1736226772 & 6.114704425 & -0.35853 \\
10 & NZD/USD & 128 & 0.989621843 & 2.060491384 & 5.462168462 & -0.224 \\
11 & EUR/JPY & 128 & 0.99206671 & 2.063575112 & 5.460526944 & -0.20097 \\
12 & EUR/CHF & 128 & 0.991928772 & 2.175657808 & 6.116291797 & -0.38479 \\
13 & EUR/GBP & 128 & 0.992315126 & 2.125032234 & 5.800669878 & -0.34398 \\
14 & USD/BRL & 128 & 0.986271542 & 1.987515102 & 5.060208482 & -0.063159 \\
15 & USD/CNH & 128 & 0.991734947 & 2.107888089 & 5.716649647 & -0.33978 \\
16 & USD/INR & 128 & 0.990322461 & 2.041044148 & 5.336929786 & -0.19754 \\
17 & USD/ZAR & 128 & 0.99075651 & 2.051054249 & 5.38480614 & -0.14857 \\
18 & USD/RUB & 128 & 0.986119801 & 2.044560147 & 5.414686924 & -0.21527 \\
19 & WTI & 128 & 0.991015688 & 2.005806129 & 5.119663516 & 0.014502 \\
20 & SOY & 128 & 0.98949516 & 1.97722104 & 4.96198781 & -0.035633 \\
21 & CORN & 128 & 0.990315723 & 1.998988571 & 5.084888371 & 0.020687 \\
22 & XAU & 128 & 0.992937334 & 2.0478933413 & 5.3498236085 & -0.14414 \\
23 & VIX & 128 & 0.986618758 & 1.980342958 & 4.996256403 & 0.029936 \\
24 & JGB & 64 & 0.9839351 & 1.9590719 & 4.8994693 & 0.0045459 \\
25 & NKY225 & 64 & 0.9801519 & 1.96546533 & 4.96602964 & -0.010206 \\
\hline
\end{tabular}
\label{table3}
\end{table}

\begin{table}[ht]
\centering
\begin{tabular}{c|ccccccccccccccccccccccccc}
\hline
 No.& 1 & 2 & 3 & 4 & 5 & 6 & 7 & 8 & 9 & 10 & 11 & 12 & 13 & 14 & 15 & 16 & 17 & 18 & 19 & 20 & 21 & 22 & 23 & 24 & 25 \\
\hline
$\mu_4$ & $\times$ &  &  &  &  &  & $\times$ &  & $\times$ &  &  & $\times$ & $\times$ &  &  &  &  &  &  &  &  &  &  &  &  \\
$\mu_6$ & $\times$ & $\times$ &  & $\times$ &  &  & $\times$ &  & $\times$ &  &  & $\times$ & $\times$ &  & $\times$ &  &  &  &  &  &  &  &  &  &  \\
\hline
\end{tabular}
\caption{Comparison to  normal i.i.d. for $\mu_4$ and $\mu_6$. The outside  of the null hypothesis  is $\times$.
The $5\%$ quantile bands are used as a reference benchmark under the Gaussian i.i.d. assumption.}
\label{T}
\end{table}

\begin{table}[tbh]
\caption{Contribution of temporal correlation of the shortest time lag in  the financial time series. The definition of temporal correlation is in Eq.(\ref{4}) and (\ref{6}).  }
\begin{center}
\begin{tabular}{|l|c|c|c|c|c|}
\multicolumn{4}{c}{}\\ \hline
Contribution of correlation/No.&1&7&9&12&13\\ 
 \hline \hline
Data&
BTC/USD	&	USD/CHF	&	USD/CAD	&	EUR/CHF   	& 	EUR/GBP\hspace{0.4cm} \\
\hline 
4 th moment $ (\%)   $&
 655.84 	&	93.47 	&	98.72	&	112.39 	&	126.18 	\\
\hline
6 th moment $ (\%)$ &
 542.99 	&	86.3	&	96.2	&	111.34 	&	122.89 	\\
\hline
\end{tabular}
\label{Corr}
\end{center}
\label{cont}
\end{table}

\section{V. Numerical simulations}
In this section,  we confirm the conclusions of  Section III using  numerical simulations.
Appendix B compares between the numerical simulations and Eq.(\ref{expmu}), (\ref{powmu}).

\subsection{A. Exponential Decay}

\begin{figure}[htbp]
\begin{center}
\begin{tabular}{cc}
 \begin{minipage}{0.5\hsize}
\begin{center}
\includegraphics[width=8cm]{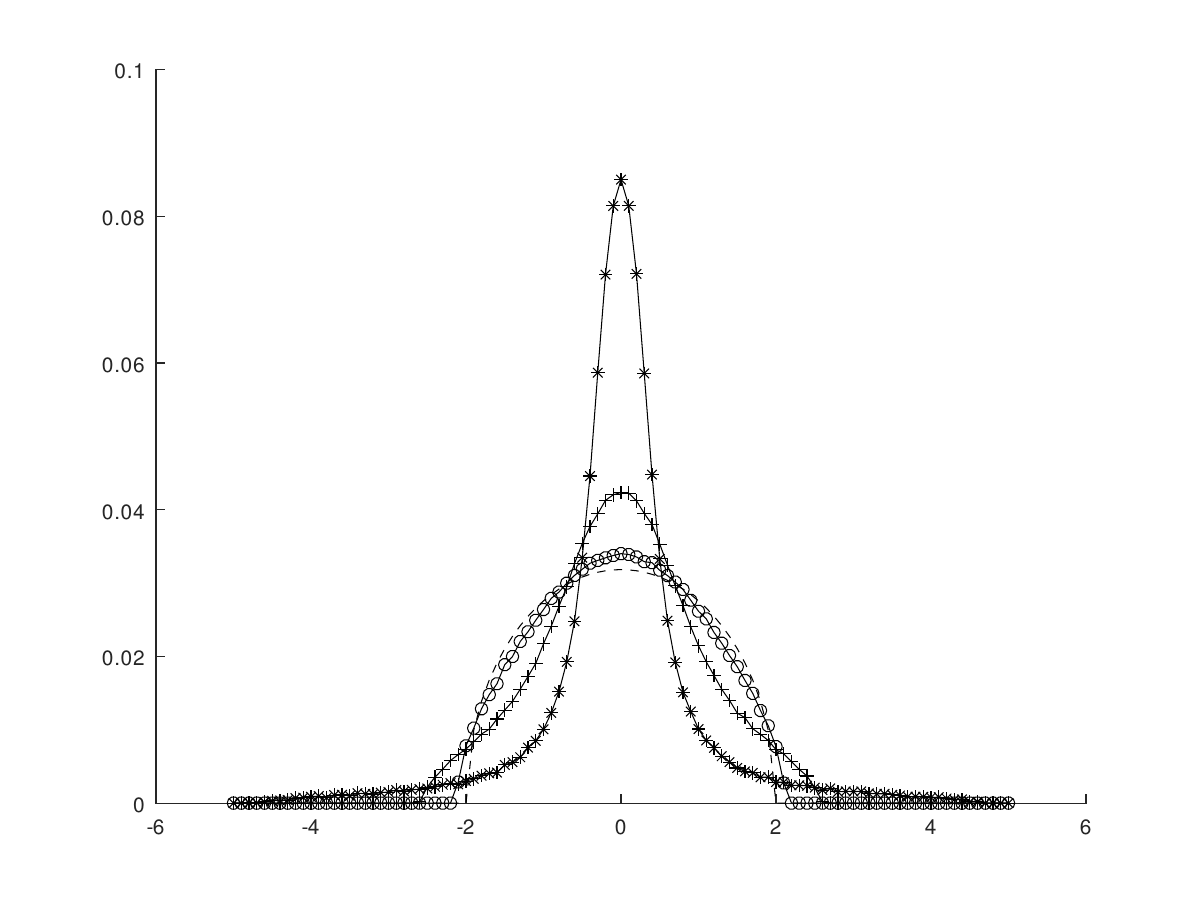}
\hspace{1.6cm} (a)
\end{center}
\end{minipage}
& 
\begin{minipage}{0.5\hsize}
\begin{center}
\includegraphics[width=8cm]{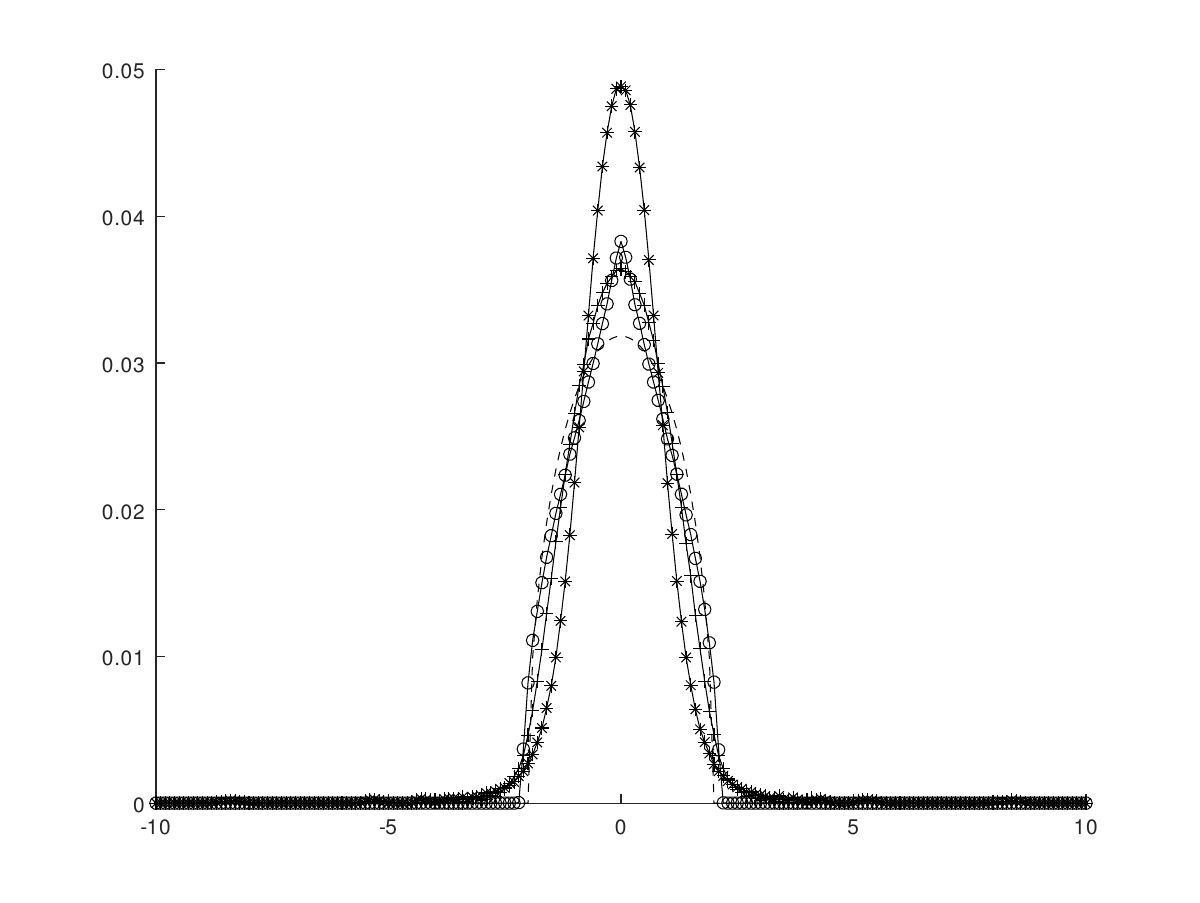}
\hspace{1.6cm} (b) 
\end{center}
\end{minipage}
\end{tabular}
\end{center}
\caption{
Plots of  the histogram  of the  deformed semicircle law:
(a)$r=0.4,0,6,0.9,$ and 
(b)$H=0.1, 0.75, 0.9.$
The dotted line represents the semicircle law.
The horizontal axis is the eigenvalue and the vertical axis 
is the frequency.
 We can confirm the fat tail  of the distribution  and the higher  peak than the semicircle law in the central for large $r$, large  $H$ and  small $H$.
}
\label{SCL}
\end{figure}

First,  we  confirm   the 
deformed  semicircle law.
We  ran 1000 simulations  for  each $r$ and 
constructed a  histogram of the eigenvalues.
The conclusions  are presented  in Fig. \ref{SCL} (a).
We can confirm  that the distributions have  a fat tail and 
 a high peak  for large $r$ and the mean of the distribution is  constant.
Note that the bulk part is different from the semicircle law for large $r$. 
In the case of eigenvalue distribution of the Wigner-L\`{e}vy matrix, the bulk part is almost same as the semicircle law for $\nu>2$, where $\nu$ is the freedom of the t distribution.
We discuss t distribution cases in Appendix C.
Conversely,   for small $r$  the distributions  are close to  the semicircle law.

\subsection{B. Fractional Brownian motion} 
Next, we confirm  the fractional Brownian motion (fBm)
 case, which has   power decay temporal correlations and  fBm is sometimes observed in the financial time series.
The relation of the indices is  as follows;
\begin{equation}
    2-2H=\gamma,
\end{equation}
where $H$ is    the Hurst index for  fBm, and $\gamma$ is the power index \cite{M}.
Hence, the transition point is  $H_c=3/4$, which corresponds 
to $\gamma_c=1/2$.

First, we confirm  the distributions of the 
deformed semicircle law for  the fBm case.
We  calculated  200 times for  each $H$ and 
created a histogram of the eigenvalues.
The conclusions are shown in Fig.\ref{SCL} (b).
We can confirm the difference from the  semicircle law in smaller $H$ 
and larger $H$. 
Conversely,  for around $H=1/2$, the distributions have    the  same shape as  the semicircle law, because $H=1/2$ is the Brownian motion.
Above $H_c=3/4$,  a  phase transition occurs and
we can observe  large eigenvalues.
In fact we can observe the several large  peaks     in Fig.\ref{SCL}, $H=0.9$.
These are the eigenvalues separated from the bulk part.

\section{VI. Phase transition of the deformed   semicircle law  }
In this section, we study the phase transition  in the case of  power decay.
We  apply  finite scaling analysis to confirm the critical exponent   and  discuss   the behavior of the   square of  scaled largest eigenvalue.

\subsection{A. Finite size scaling}

In this subsection, we consider  the  finite size scaling.
The scaling function  is introduced to  the scaled squared largest eigenvalue, 
\begin{equation}
\frac{x_1^2}{N^{1+\gamma/\nu}}=f(N^{1/\nu} t),
    \label{sc}
\end{equation}
where
$t=(H-H_c)/H_c$ and $f(x)$ is a scaling function.
It is  hypothesized that  the data of several $N$ are on  the curve, Eq.(\ref{sc}).  
We assume that  the scaling function is $f(x)\sim 0$  in the limit $x\rightarrow -\infty$ and $f(x)\sim O(x)$   in the limit $x \rightarrow \infty$.

This scaling  hypothesis  is confirmed in Fig \ref{scaling}, where we set  $\gamma=-1$ and $1/\nu=0.65$.
The exponent was estimated phenomenologically by visual curve collapse.
No optimization procedure was employed in this estimation.
In the  MPD case $\gamma/\nu=-0.75$ \cite{Hisakado5}.

In these hypotheses, the following  relation can be obtained,
\begin{eqnarray}
m\equiv \frac{ x_1^2 }{ N}&\propto& t=(H-H_c)/H_c \hspace{2cm} H-H_c>>N^{-1/\nu}
\nonumber  \\ 
&\propto& 0. \hspace{5cm} H_c-H>>N^{-1/\nu}
\label{sc2}
\end{eqnarray}
Therefore, the order parameter is $m=x_1^2 /N$, the scaled squared largest eigenvalue.

\begin{figure}[htbp]
\begin{center}
\includegraphics[width=8cm]{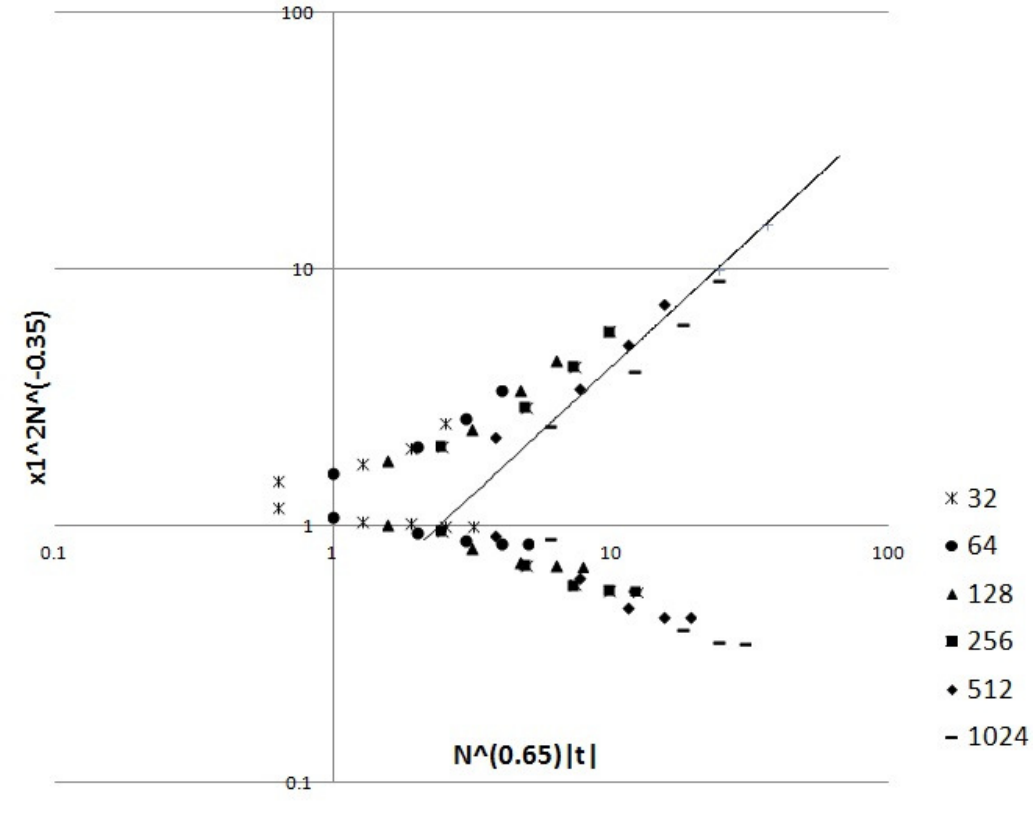}
\end{center}
\caption{
Figure  shows    the finite size scaling plot  for  fBm.
The horizontal axis is  a log  of $ N^{0.65}|t|=N^{0.65}|H-H_c|/H_c$, RHS of  Eq.(\ref{sc}) and 
the vertical axis is a log  of $x_1^2 N^{-0.35}$, LHS of Eq.(\ref{sc}).
Here, we use $H_c=3/4$.
We can  confirm that  $N=32, 64,128, 256, 512, 1024$ are on a curve  near the critical point and we show  the asymptotic line, $x$, which corresponds to the scaling function.
The solid line represents  the asymptotic line  $t$ and the trend is $\beta=1$,  which is consistent with Eq.(\ref{sc2}).
}
\label{TSP}
\end{figure}

The slope of the asymptotic line in Fig \ref{TSP}  is $\beta=1$  for large $t$,
 and it is consistent with   Eq.(\ref{sc}) and  Eq.(\ref{sc2}).
Therefore, the critical exponent  is  $\beta=1$ and it is  same as the MPD case.

We obtain the correlation  length,   $\xi_{\infty}$
\begin{equation}
    \xi_{\infty}\propto t^{-\nu}=t^{ -1/0.65} \hspace{0.8cm}{\rm at} \hspace{2cm}H\sim H_c,
\end{equation}
where  $\xi_{\infty}$ is the correlation length.
In the case of MPD, we obtained $1/\nu=0.75$ which is different from this model \cite{Hisakado5}.
The scaling exponents were estimated phenomenologically from visual curve collapse and should therefore be regarded as approximate rather than precise critical estimates. 
Nevertheless, the collapse remained reasonably stable within the ranges 0.72–0.76 for the MP case and 0.60–0.67 for the semicircle-law case.
The difference is from the difference in  the symmetry of the matrix.

\subsection{B. Scaled squared largest eigenvalue} 
\begin{figure}[htbp]
\begin{center}
\begin{tabular}{cc}
 \begin{minipage}{0.5\hsize}
\begin{center}
\includegraphics[width=8cm]{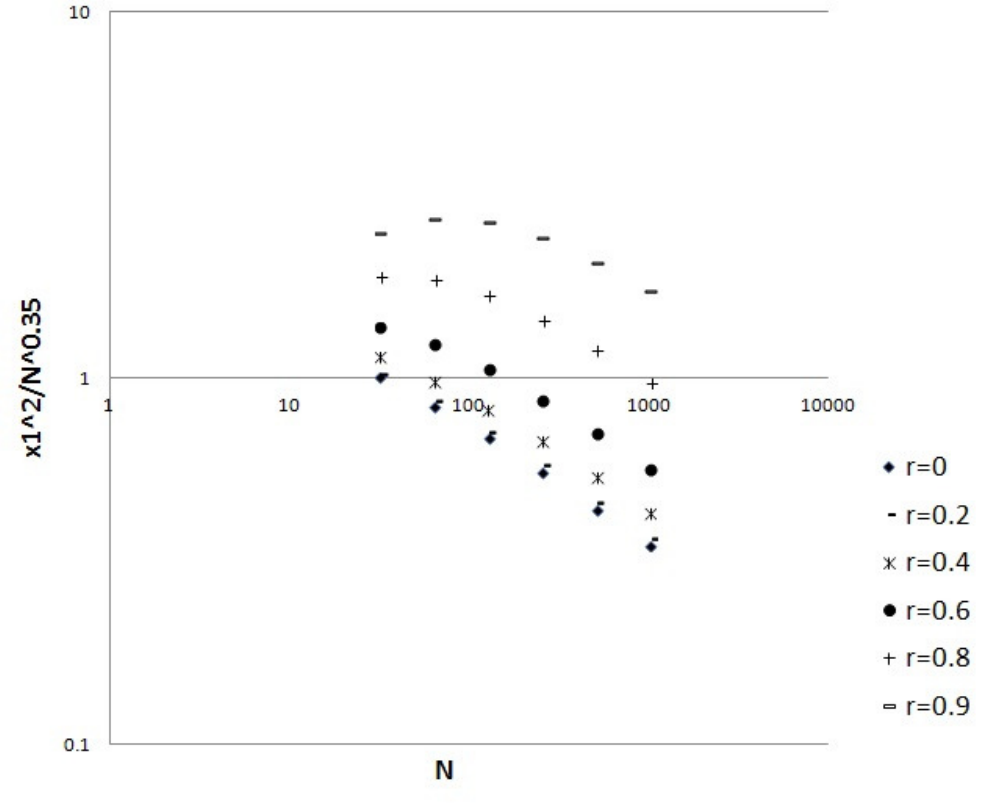}
\hspace{1.6cm} (a)
\end{center}
\end{minipage}
& 
\begin{minipage}{0.5\hsize}
\begin{center}
\includegraphics[width=9.5cm]{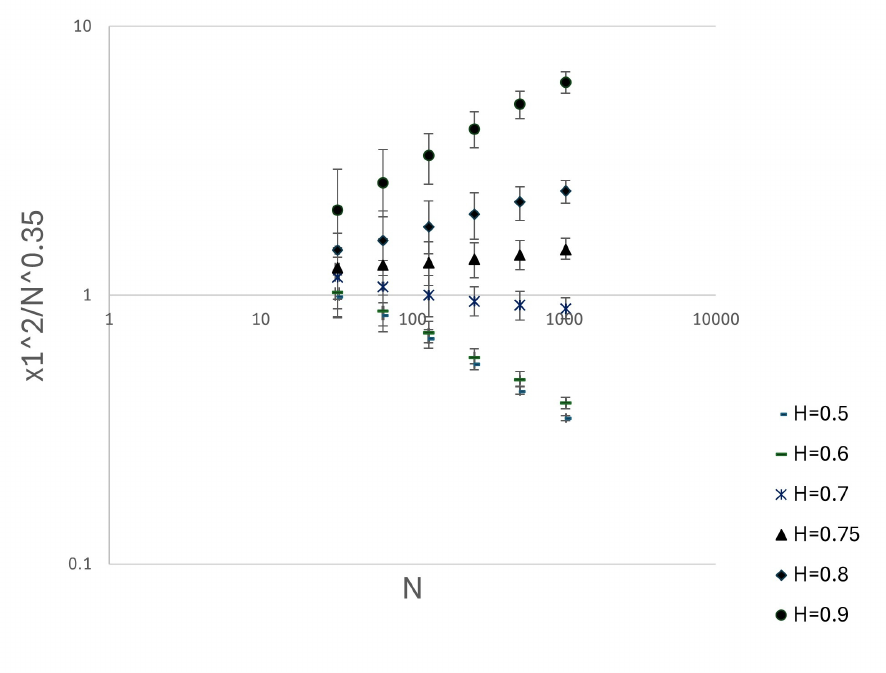}
\hspace{1.6cm} (b) 
\end{center}
\end{minipage}
\end{tabular}
\end{center}
\caption{
Figures (a)  and (b)  show    the scaled squared  largest eigenvalue, $x_1^2/N^{0.35}$,  for exponential decay and fBm.
The horizontal axis is a  log  of the matrix size $N$,  and the vertical axis is  a log  of the  scaled largest eigenvalue $x_1^2/N$. The 5\% and 95\% quantile bands of (b) are obtained from numerical simulations.
}
\label{scaling}
\end{figure}

In this subsection, we confirm 
the scaled squared largest eigenvalue, $0<x_1^2/N^{0.35}\leq 1$ which is LHS of Eq.(\ref{sc}). 
In Fig.\ref{scaling}  we show the scaled squared eigenvalue for (a)  exponential  decay and (b) fBm cases.
 The horizontal axis is    a log plot of  matrix size $N$, and the vertical axis is  a log plot of the scaled squared largest eigenvalue $x_1^2/N^{0.35}$.
 In the exponential decay case,   the scaled squared largest eigenvalue decreases to 0 as Fig.\ref{scaling} (a), because $x_1$ is finite.
 In the case of fBm,
 we confirm the difference in $H>H_c$ and $H<H_c$.
 In $H<H_c$  the scaled squared largest eigenvalue decreases to $0$ when $H<H_c$.
 Conversely, the scaled squared largest eigenvalue increases when $H>H_C$.
When $H<H_c$, $x_1$ is finite  and when $H>H_c$, $x_1$ is infinite.

\section{VII. Concluding Remarks}

In this paper, we have investigated the spectral properties of Wigner-type random matrices constructed from time series with temporal correlations. We have shown that such correlations deform the semicircle law and induce a phase transition in the case of power-law decay.
Our analysis is based on moment calculations and numerical observations, rather than a rigorous proof of convergence.

A central result of this work is that the mechanism of the transition differs qualitatively from that observed in the Wishart ensemble. In particular, while the transition in the Wishart case is governed by the second moment, the present system is controlled by the fourth moment, which increases with the strength of correlations. Furthermore, the finite-size scaling behavior is characterized by a critical exponent that differs from that of the Wishart ensemble.
We show the comparisons in Table.\ref{comp}.

These findings indicate that correlation-induced spectral transitions are not universal across random matrix ensembles, but instead depend on the structural properties of the matrices. In this sense, the Wigner and Wishart ensembles
belong to distinct   universality classes with respect to correlation effects.

Our results are supported by numerical simulations based on both synthetic data, including fractional Brownian motion, and empirical financial time series. While finite-size effects and higher-order moment fluctuations remain important near the critical regime, the overall behavior is consistent with the proposed framework.
In fact, we have applied the framework to financial time series and   confirmed  that certain foreign exchange rates deviate significantly from the semicircle law, consistent with their stronger temporal dependencies.

These results open the possibility of classifying correlation-induced spectral transitions in terms of universality classes, depending on the underlying matrix structure. Further theoretical analysis, including approaches based on the Stieltjes transform, as well as more robust statistical characterizations, remain important directions for future work.

\begin{table}[h]
\centering
\caption{Comparison between the Marchenko--Pastur law (Wishart ensemble) 
and the semicircle law (Wigner ensemble) under correlation-induced transitions. 
Both ensembles share the same critical point $\gamma_c = 1/2$, but differ in 
the moment order governing the transition, the combinatorial coefficient of 
the correlation-induced correction, and the critical exponent $1/\nu$. 
The differences in combinatorial coefficients ($2$ vs.\ $4/3$)  and governing moments (2nd  and 4th ) reflect the 
distinct matrix construction principles and is conjectured to contribute to 
the observed difference in critical exponents.}
\label{tab:comparison}
\begin{tabular}{lcc}
\hline\hline
 & Marchenko--Pastur (Wishart) & Semicircle (Wigner) \\
\hline
Matrix type          & $X^TX$ (sample covariance)      & Symmetric random matrix        \\
Limiting law         & Marchenko--Pastur distribution  & Semicircle distribution        \\
With correlations    & Deformation of MP law           & Deformation of semicircle law  \\
Governing moment     & 2nd moment                      & 4th moment                     \\
Combinatorial coefficient & $2$                        & $4/3$ (Appendix A.)                         \\
Critical point $\gamma_c$ & $1/2$                      & $1/2$                          \\
Critical exponent $1/\nu$ & 0.75(0.72$\sim$ 0.76)                    & 0.65   (0.60$\sim$0.67)                      \\
\hline\hline
\end{tabular}
\label{comp}
\end{table}

\def\thesection{Appendix \Alph{section}}

\section{Appendix A  the calculation of second moment}

 If the selected term crosses the diagonal elements, the correlation becomes zero.
 Therefore, we calculate the  ratio of pairs that cross the diagonal element.
\begin{equation}
\lim_{N\rightarrow \infty} \frac{1}{N^3}\sum_{k=1}^{N-1}k(N-k)=\frac{1}{6}.
\end{equation}
The normalization  of all pairs is
\begin{equation}
\lim_{N\rightarrow \infty} \frac{1}{N^3}   {}_ N C_2 N=\frac{1}{2}.
\end{equation}
Therefor, the ratio of correlated pairs is $1-1/6/1/2=2/3$ 
\begin{eqnarray}
    \mu_4&=&\frac{1}{N^3}\sum_{j=1}^N\sum_{l 
 =1}^N\sum_{m =1}^N
    \sum_{n =1}^N< S_{j l} S_{l m}S_{m n} S_{n j}> \nonumber \\
    &=&
    2+\frac{2\cdot2}{3} \sum_{i=1} d_i^2=2+\frac{4}{3} \sum_{i=1} d_i^2,
\label{mu1}
\end{eqnarray}

\section{Appendix B Comparison of the  fourth and sixth  moments }

In this Appendix we compare  between theory and numerical simulations in Table \ref{Com1} and  Table \ref{Com2}.
We use Eq.(\ref{expmu}),   Eq.(\ref{expmu2}),  and Eq.(\ref{powmu}) as   the theories.
The uncertainly ranges are obtained 1000 and 200 times simulations for exponential decay  and fBm cases, respectively.
The difference for the exponential decay case  is inside of the bands  in Table \ref{Com1}.
The discrepancies  observed at small $H$ (e.g., $H = 0.1, 0.2, 0.3$) and large $H$ 
(e.g., $H = 0.7$) in Table~\ref{Com2}, where the theoretical 
values exceed the simulation averages, are attributable to finite-size effects. 
Specifically, the theoretical values in Eq.(\ref{powmu}) are derived 
in the limit $N \to \infty$, whereas the simulations are performed at finite 
$N$. The finite-size correction scales  of the difference  are  $\sim N^{1-2\gamma}$, where $\gamma>1/2$. 
The simulation averages are expected to approach the theoretical values 
as $N$ increases.

\begin{table}[h]
\centering
\caption{Comparison of $\mu_4$ and $\mu_6$ in exponential decay case ($N=1024$). 
The 5\% and 95\% quantile bands are obtained from numerical simulations.}
\label{tab:exp_moments}
\begin{tabular}{ccccccccc}
\hline\hline
 & \multicolumn{4}{c}{4th moment $\mu_4$} & \multicolumn{4}{c}{6th moment $\mu_6$} \\
\cmidrule(lr){2-5}\cmidrule(lr){6-9}
$r$ & Theory & Average & 5\% & 95\% & Theory & Average & 5\% & 95\% \\
\hline
$0.0$ & $2$       & $2.0008$ & $1.9880$ & $2.0141$ & $5$          & $5.0061$  & $4.9518$  & $5.0592$  \\
$0.1$ & $2.01347$ & $2.0149$ & $2.0012$ & $2.0290$ & $5.08108$    & $5.0898$  & $5.0318$  & $5.1484$  \\
$0.2$ & $2.0556$  & $2.0563$ & $2.0409$ & $2.0712$ & $5.33796$    & $5.3452$  & $5.2787$  & $5.4110$  \\
$0.3$ & $2.1319$  & $2.1323$ & $2.1160$ & $2.1486$ & $5.81729$    & $5.8257$  & $5.7509$  & $5.9046$  \\
$0.4$ & $2.2540$  & $2.2552$ & $2.2360$ & $2.2756$ & $6.62056$    & $6.6427$  & $6.5464$  & $6.7428$  \\
$0.5$ & $2.4444$  & $2.4452$ & $2.4196$ & $2.4705$ & $7.96296$    & $8.0088$  & $7.8650$  & $8.1525$  \\
$0.6$ & $2.7500$  & $2.7494$ & $2.7157$ & $2.7819$ & $10.34375$   & $10.4488$ & $10.2184$ & $10.6738$ \\
$0.7$ & $3.2810$  & $3.2799$ & $3.2301$ & $3.3290$ & $15.14789$   & $15.4488$ & $15.0374$ & $15.8829$ \\
$0.8$ & $4.3703$  & $4.3604$ & $4.2746$ & $4.4476$ & $27.65021$   & $28.5731$ & $27.5169$ & $29.6178$ \\
$0.9$ & $7.6842$  & $7.6123$ & $7.3716$ & $7.8506$ & $87.57064$   & $91.7924$ & $86.3279$ & $97.2030$ \\
\hline\hline
\end{tabular}
\label{Com1}
\end{table}

\begin{table}[h]
\centering
\caption{Comparison of $\mu_4$ in power decay (fBm) case ($N=1024$). 
The 5\% and 95\% quantile bands are obtained from numerical simulations. 
N.A.\ indicates that the theoretical value diverges for $H > H_c = 3/4$.}
\label{tab:fBm_mu4}
\begin{tabular}{ccccc}
\hline\hline
$H$ & Theory & Average & 5\% & 95\% \\
\hline
$0.10$ & $2.2428$ & $2.1887$ & $2.1831$ & $2.1953$ \\
$0.20$ & $2.1581$ & $2.1321$ & $2.1270$ & $2.1378$ \\
$0.30$ & $2.0835$ & $2.0743$ & $2.0694$ & $2.0790$ \\
$0.40$ & $2.0259$ & $2.0246$ & $2.0194$ & $2.0293$ \\
$0.50$ & $2$      & $2.0014$ & $1.9964$ & $2.0056$ \\
$0.60$ & $2.054$  & $2.0521$ & $2.0466$ & $2.0579$ \\
$0.70$ & $2.493$  & $2.4018$ & $2.3781$ & $2.4282$ \\
$0.75$ & N.A.     & $2.9893$ & $2.9202$ & $3.0640$ \\
$0.80$ & N.A.     & $4.3839$ & $4.1258$ & $4.5979$ \\
$0.90$ & N.A.     & $14.4419$ & $13.2015$ & $15.9914$ \\
\hline\hline
\end{tabular}
\label{Com2}
\end{table}

\section{Appendix C. t distribution (Supplementary comparison) }

\begin{figure}[htbp]
\begin{tabular}{ccc}
\begin{minipage}{0.33\hsize}
\begin{center}
 \includegraphics[width=5.cm]{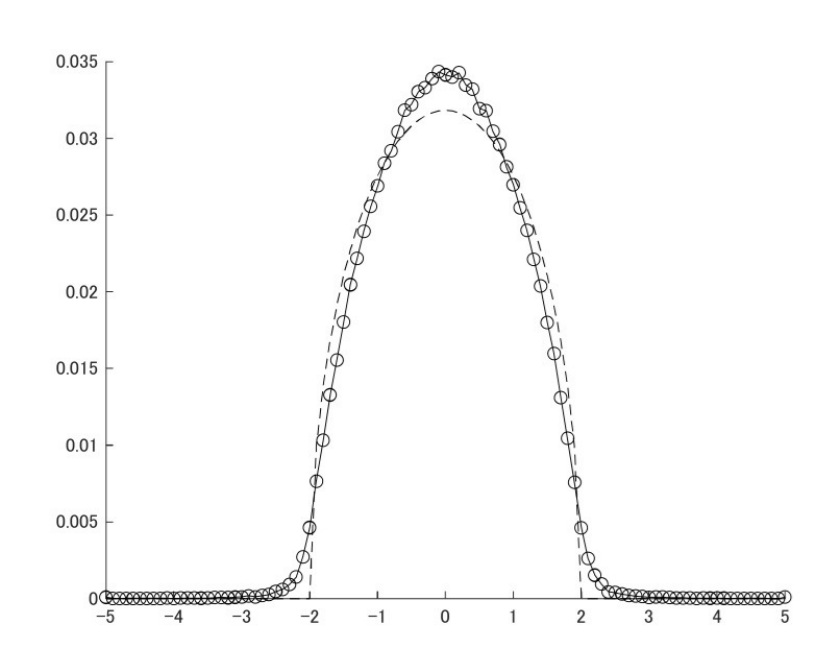} 
\hspace{1.6cm} (a) $\nu=3.5$
\end{center}
\end{minipage}
 &
\begin{minipage}{0.33\hsize}
\begin{center}
 \includegraphics[width=5.cm]{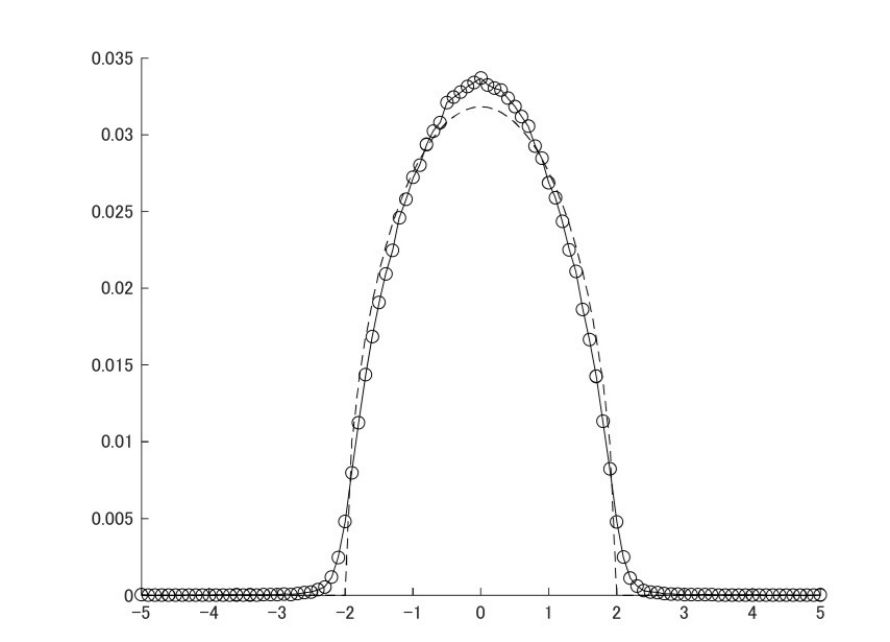} 
\hspace{1.6cm} (b) $\nu=4.0$
\end{center}
\end{minipage}
 &
 \begin{minipage}{0.33\hsize}
\begin{center}
 \includegraphics[width=5.cm]{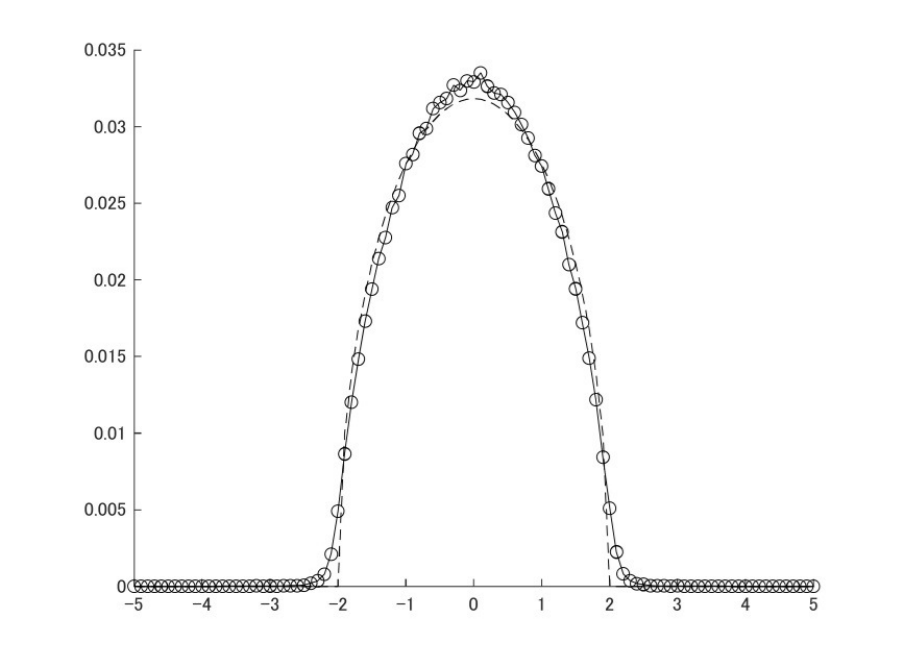} 
\hspace{1.6cm} (c) $\nu=4.5$
\end{center}
\end{minipage} 
\end{tabular}  
\caption{Plots of the histogram of  eigenvalue distribution
(a) $\nu=3.5$, (b) $\nu=4.0$, (c) $\nu=4.5$.
The horizontal axis represents  the eigenvalues and the vertical axis 
represents the frequency.
The real line presents the distribution  of the time series  and the dotted line represents  the semicircle law. 
We can confirm the bulk part is almost semi-circle law and the edge is different from $2$ when $\nu<4$.
}
\label{MPD4}
\end{figure}

\begin{figure}[htbp]
\begin{center}
\includegraphics[width=8cm]{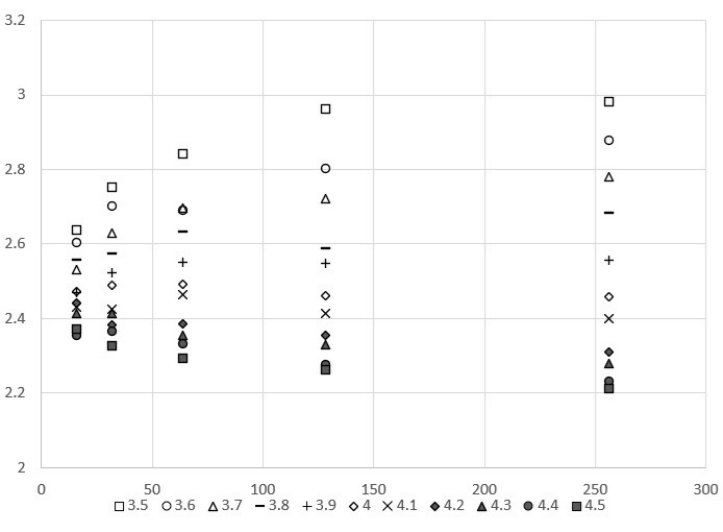}
\end{center}
\caption{Convergence to the maximum eigenvalue. The horizontal axis represents  the matrix size, $N$ and the vertical axis 
represents the 95\% point of the maximum eigenvalue.
When $\nu=4$, the the 95\% point of the maximum eigenvalue converges. This is the conclusion of 5000 trials in each  $N$.
}
\label{con}
\end{figure}

In this Appendix we consider the t distributed  elements of  Wigner matrix.
When the degree of  freedom $\nu$ is below 4,   the fourth  moment  of t distribution  is infinite.
When the  degree of freedom $\nu$ is above  4,   the fourth  moment of t distribution  is finite.
Here we consider the case of no temporal correlation.
We can observe the similar phenomena in the section III.
The fourth moment  of the eigenvalue distribution is $\mu_4=2$ when $\nu >4$ and
 $\nu\leq 4$,  $\mu_4$ goes infinity.
However, this is not a   phase transition.
We show the plot of the eigenvalue distribution in Fig.\ref{MPD4}.
When $\nu>4$, the distribution is almost same as the  semicircle law and the maximum eigenvalue converges to $2$ as the matrix size $N$  increases.
When  $\nu\leq 4$  the  maximum eigenvalue  does not  converge to $2$, but increases.
In Fig.\ref{con} we show the convergence of the 95\% point of maximum eigenvalue.
As  the matrix size, $N$ increases, it converges to 2. 
We can confirm the bulk part is almost semi-circle law and the edge is different from $2$ when $\nu<4$ \cite{B1,B2}.
When $\nu>>4$,  we can confirm the semicircle law \cite{tao, tao2}.

\section{Appendix D. Bootstrap for the financial time series}

In this Appendix, we evaluate the statistical uncertainties of the estimates reported in Table 2. The confidence intervals were obtained using the bootstrap method. Specifically, for each dataset, bootstrap method  was repeatedly performed, and the $5\%$ and $95\%$ percentile values were calculated to estimate the uncertainty ranges of $\mu_4$ and $\mu_6$. The numbers of bootstrap samples and data sizes used in the analysis are also listed in the table.
The bootstrap samples were generated using overlapping windows (sampling with overlap) in order to preserve the temporal correlation structure of the original time series as much as possible.

\begin{table*}[t]
\centering
\caption{Bootstrap confidence intervals for $\mu_4$ and $\mu_6$ for various financial assets.}
\begin{tabular}{c l c c c c c c}
\hline
No & Data & Bootstrap & $N$ & $\mu_4$ 5\% & $\mu_4$ 95\% & $\mu_6$ 5\% & $\mu_6$ 95\% \\
\hline
1  & BTC/USD & 1000 & 512 & 1.972695 & 2.079119 & 5.011060 & 5.437309 \\
2  & XRP/USD & 800  & 256 & 1.951266 & 2.132678 & 4.916165 & 5.681266 \\
3  & ETH/USD & 1000 & 128 & 1.770586 & 2.245336 & 4.232598 & 6.161839 \\
4  & EUR/USD & 1000 & 128 & 1.985531 & 2.195846 & 5.104403 & 6.101104 \\
5  & USD/JPY & 1000 & 128 & 1.974937 & 2.143215 & 5.077340 & 5.859002 \\
6  & GBP/USD & 1000 & 128 & 1.901383 & 2.209596 & 4.769590 & 6.106649 \\
7  & USD/CHF & 1000 & 128 & 2.038203 & 2.206354 & 5.431083 & 6.324475 \\
8  & AUD/USD & 1000 & 128 & 1.931659 & 2.170240 & 4.918608 & 5.977065 \\
9  & USD/CAD & 1000 & 128 & 2.018599 & 2.364635 & 5.345848 & 7.094168 \\
10 & NZD/USD & 1000 & 128 & 1.977178 & 2.142831 & 5.074376 & 5.859172 \\
11 & EUR/JPY & 1000 & 128 & 1.992023 & 2.130719 & 5.119283 & 5.812211 \\
12 & EUR/CHF & 1000 & 128 & 2.095441 & 2.254896 & 5.701272 & 6.546380 \\
13 & EUR/GBP & 1000 & 128 & 2.049091 & 2.217592 & 5.422231 & 6.265321 \\
14 & USD/BRL & 1000 & 128 & 1.856801 & 2.108957 & 4.548892 & 5.564892 \\
15 & USD/CNH & 1000 & 128 & 2.015465 & 2.214615 & 5.306589 & 6.170781 \\
16 & USD/INR & 1000 & 128 & 1.886790 & 2.216207 & 4.683206 & 6.077910 \\
17 & USD/ZAR & 1000 & 128 & 1.948590 & 2.195704 & 4.954848 & 6.003061 \\
18 & USD/RUB & 1000 & 128 & 1.838772 & 2.278320 & 4.587273 & 6.464053 \\
19 & WTI     & 1000 & 128 & 1.926311 & 2.082001 & 4.777799 & 5.480785 \\
20 & SOY     & 1000 & 128 & 1.866640 & 2.080680 & 4.536553 & 5.389560 \\
21 & CORN    & 1000 & 128 & 1.842046 & 2.181902 & 4.475750 & 5.810267 \\
22 & XAU     & 1000 & 128 & 1.935784 & 2.170339 & 4.895872 & 5.888762 \\
23 & VIX     & 1000 & 128 & 1.824046 & 2.102088 & 4.332645 & 5.509567 \\
24 & JGB     & 200  & 64  & 1.827019 & 2.095425 & 4.333298 & 5.467612 \\
25 & NKY225  & 200  & 64  & 1.781749 & 2.214389 & 4.261933 & 5.944136 \\
\hline
\end{tabular}
\label{tab:bootstrap_financial}
\end{table*}

\end{document}